%% file: wznnlo.tex
\documentclass[12pt]{article}
\usepackage[utf8]{inputenc}
\usepackage{epsf,amsmath,amssymb,graphicx,dcolumn}
\usepackage{caption}
\usepackage{subcaption}
\usepackage{scalefnt,ulem,pstricks}
\usepackage{booktabs,multirow,tabularx}
\usepackage{cite,hyperref}
\usepackage{cleveref}
\usepackage{color}
\usepackage{rotating}
\usepackage{microtype}

\providecommand{\href}[2]{#2}

\newcommand\as{\alpha_{\mathrm{S}}}

\def\to{\rightarrow}

\def\qT{q_T}

\def\GeV{\mathrm{GeV}}
\def\TeV{\mathrm{TeV}}

\newcommand\Matrix{{\sc Matrix}}
\newcommand\Munich{{\sc Munich}}
\newcommand\OpenLoops{{\sc OpenLoops}}
\newcommand\Collier{{\sc Collier}}
\newcommand{\CutTools}{{\sc CutTools}}
\newcommand{\OneLOop}{{\sc OneLOop}}

\newcommand{\zz}{\ensuremath{ZZ}}
\newcommand{\ww}{\ensuremath{W^+W^-}}
\newcommand{\wz}{\ensuremath{W^\pm Z}}
\newcommand{\wpz}{\ensuremath{W^+Z}}
\newcommand{\wmz}{\ensuremath{W^-Z}}
\newcommand{\z}{\ensuremath{Z}}
\newcommand{\w}{\ensuremath{W}}

\newcommand{\abbrev}{}

\newcommand{\nnll}{\text{\abbrev NNLL}}
\newcommand{\lo}{\text{\abbrev LO}}
\newcommand{\nlo}{\text{\abbrev NLO}}
\newcommand{\nnlo}{\text{\abbrev NNLO}}

\newcommand{\qcd}{{\abbrev QCD}}

\newcommand{\cme}{centre-of-mass energy}
\newcommand{\cmes}{centre-of-mass energies}

\newcommand{\citere}[1]{Ref.~\cite{#1}}
\newcommand{\citeres}[1]{Refs.~\cite{#1}}
\newcommand{\reffi}[1]{Fig.~\ref{#1}}

\newcommand{\refta}[1]{Table~\ref{#1}}

\newcommand\Tstrut{\rule{0pt}{3.0ex}}         
\newcommand\Bstrut{\rule[-1.5ex]{0pt}{0pt}}   

\begin{document} 
\begin{titlepage}
\renewcommand{\thefootnote}{\fnsymbol{footnote}}
\begin{flushright}
ZU-TH 15/16\\
MITP 16-037\\
NSF-KITP-16-046\\
DESY 16-074
\end{flushright}
\vspace*{2cm}

\begin{center}
{\Large \bf \wz{} production at hadron colliders in \nnlo{} \qcd{}}
\end{center}

\par \vspace{2mm}
\begin{center}
{\bf Massimiliano Grazzini$^{(a)}$},
{\bf Stefan Kallweit$^{(b,c)}$},\\[0.2cm] {\bf Dirk Rathlev$^{(d)}$} and {\bf Marius Wiesemann$^{(a)}$}
\vspace{5mm}

$^{(a)}$Physik-Institut, Universit\"at Z\"urich, CH-8057 Z\"urich, Switzerland 

$^{(b)}$PRISMA Cluster of Excellence, Institute of Physics,\\[0.1cm]
Johannes Gutenberg University, D-55099 Mainz, Germany

$^{(c)}$Kavli Institute for Theoretical Physics,\\[0.1cm]
University of California, Santa Barbara, CA 93106, USA

$^{(d)}$Theory Group, Deutsches Elektronen-Synchrotron, D-22607 Hamburg, Germany 

\vspace{5mm}

\end{center}

\par \vspace{2mm}
\begin{center} {\large \bf Abstract} \end{center}
\begin{quote}
\pretolerance 10000

We report on the first computation of the next-to-next-to-leading order (\nnlo{})
QCD corrections to \wz{} production in proton collisions.
We consider both the inclusive production of on-shell \wz{} pairs
at LHC energies
and the total \wz{} rates including off-shell effects of the $W$ and $Z$ bosons.
In the off-shell computation, the invariant mass of the lepton pairs from the 
$Z$ boson decay is required to be in a given mass window, and
the results are compared with the corresponding measurements obtained by the 
ATLAS and CMS collaborations.
The \nnlo{} corrections range from $8\%$ at $\sqrt{s}=7$\,TeV 
to $11\%$ at $\sqrt{s}=14$\,TeV
and significantly improve the agreement with the LHC data at $\sqrt{s}=7$ and $8$\,TeV.

\end{quote}

\vspace*{\fill}
\begin{flushleft}
April 2016

\end{flushleft}
\end{titlepage}

\setcounter{footnote}{1}
\renewcommand{\thefootnote}{\fnsymbol{footnote}}

The production of \wz{} pairs at the Large Hadron Collider~(LHC)
provides an important test of the electroweak~(EW) sector of
the Standard Model~(SM) at the TeV scale.
Due to its sensitivity to the gauge-boson self-interactions,
small deviations in the observed rates or in the kinematical distributions could give 
a hint of new physics. Such effects can be modelled on the basis of 
higher-dimensional operators in the form of anomalous couplings. 
With the increasing reach in energy of LHC Run 2, \wz{} measurements 
will be a powerful tool to extend the current bounds on these effective couplings.
The \wz{} process also constitutes an irreducible background in many new-physics searches, 
see for example \citere{Morrissey:2009tf}.

With its relatively small cross section \wz{} production yielded only a limited number
of events at the Tevatron~\cite{Lipeles:2007ky,Abazov:2012cj},
but its cross section has been measured with good precision at the LHC by both 
ATLAS~\cite{Aad:2012twa,Aad:2016ett} and CMS~\cite{CMS:2013qea}
at \cmes{} of $7$ and $8$\,TeV. An early measurement of the \wz{} cross section 
at $13$\,TeV by CMS is also already available~\cite{CMS:2016kma}.

On the theory side, the first next-to-leading order (\nlo{}) predictions for on-shell 
\wz{} production were obtained long ago~\cite{Ohnemus:1991gb}. 
Leptonic decays were added in \citere{Ohnemus:1994ff}, initially neglecting spin 
correlations in the virtual matrix elements.
The computation of the relevant one-loop helicity amplitudes~\cite{Dixon:1998py} enabled
the first complete \nlo{} calculations~\cite{Campbell:1999ah,Dixon:1999di,Campbell:2011bn}, including spin correlations and off-shell effects.
The \nlo{} \qcd{} corrections to off-shell $\wz{}+{\rm jet}$ production were presented in 
\citere{Campanario:2010hp}, 
and the on-shell EW corrections to \wz{} production in \citere{Bierweiler:2013dja}.

While the \ww{} and \zz{} cross sections can be computed at \nnlo{} in 
the on-shell approximation using two-loop amplitudes for two massive vector bosons 
of the same mass, as done in \citere{Cascioli:2014yka} and \citere{Gehrmann:2014fva}, 
respectively, the \wz{} production process requires the amplitudes with different 
masses of the vector-boson pairs already in the on-shell approximation. 
The required two-loop amplitudes were presented in 
\citeres{Caola:2014iua,Gehrmann:2015ora} in the form of helicity amplitudes for all 
vector-boson pair production processes, 
enabling at the same time the computation of \nnlo{} corrections to \wz{} production 
as well as the inclusion of off-shell effects and the leptonic decays of the vector bosons 
at the \nnlo{} level. In the meantime, the implementation
of the two-loop form factors for the helicity amplitudes into a numerical code 
provided by the authors of \citere{Gehrmann:2015ora} has been used to obtain \nnlo{} 
predictions for the $\zz\to 4\ell$ process in \citere{Grazzini:2015hta} and 
the $\ww{}\to2\ell2\nu$ process in \citere{wwnnlopaper}.

\wz{} production is the only remaining diboson process for which a 
complete \nnlo{} calculation was not available so far.
In this letter, we finally close this gap by providing \nnlo{} predictions for 
the \wz{} cross section at various LHC energies, 
which thus represents an important milestone in the \nnlo{} programme.
We restrict ourselves to presenting inclusive results, both for on-shell \wz{} production, 
and including all off-shell effects, 
but applying minimal mass cuts on the reconstructed \z{} boson. 
Our off-shell calculation in particular includes
the singly-resonant contributions of the form $pp\to W^\pm\to 3\ell\nu$, 
the resonant $W^\pm\gamma^*$ contributions and all interference terms.
The computation presented here thus paves the way to a fully-differential computation 
of the process $pp\to \ell^{(')\pm} \nu_{\ell^{(')}} \ell^+\ell^-$ in the future.

Our calculation is performed with the numerical program
\Matrix\footnote{\Matrix{} is the abbreviation of 
``\Munich{} Automates qT subtraction and Resummation
to Integrate Cross Sections'', by M.~Grazzini, S.~Kallweit, D.~Rathlev, M.~Wiesemann. 
In preparation.}, which combines the $\qT$-subtraction~\cite{Catani:2007vq} and 
-resummation~\cite{Bozzi:2005wk} formalisms with the 
\Munich{} Monte-Carlo framework~\cite{Kallweit:Munich}.
\Munich{} provides a fully-automated implementation of the Catani--Seymour dipole 
subtraction method~\cite{Catani:1996jh,Catani:1996vz},
an efficient phase-space integration,
as well as an interface to the one-loop generator \OpenLoops{}~\cite{Cascioli:2011va} 
to obtain all required (spin- and colour-correlated) 
tree-level and one-loop amplitudes.
For the numerically stable evaluation of tensor integrals, \OpenLoops{} relies on 
the \Collier{} library~\cite{Denner:2014gla,Denner:2016kdg}, which is based on the 
Denner--Dittmaier reduction techniques~\cite{Denner:2002ii,Denner:2005nn} and the scalar 
integrals of~\citere{Denner:2010tr}.
To deal with problematic phase-space points, a rescue system is provided,
which employs the quadruple-precision implementation of the OPP method in 
\CutTools{}~\cite{Ossola:2007ax} and scalar integrals from 
\OneLOop{}~\cite{vanHameren:2010cp}.
Our implementation of $\qT$ subtraction and resummation\footnote{The first application 
of the transverse-momentum resummation framework implemented in \Matrix{} at 
\nnll{}+\nnlo{} to on-shell \ww{} and \zz{} production was presented 
in \citere{Grazzini:2015wpa} (see also \citere{Wiesemann:2016tae} for more details).}
for the production of colourless final states is fully general, 
and it is based on the universality of the hard-collinear 
coefficients~\cite{Catani:2013tia} appearing in 
the transverse-momentum resummation formalism.
These coefficients were explicitly computed for quark-initiated processes
in \citeres{Catani:2012qa,Gehrmann:2012ze,Gehrmann:2014yya}.
For the two-loop helicity amplitudes we use the results of \citere{Gehrmann:2015ora},
and of \citere{Matsuura:1988sm} for Drell-Yan like topologies.

A preliminary version of \Matrix{} has been employed in the \nnlo{} computations of
\citeres{Grazzini:2013bna,Cascioli:2014yka,Gehrmann:2014fva,Grazzini:2015nwa,%
Grazzini:2015hta,wwnnlopaper}, 
and in the resummed calculation of \citere{Grazzini:2015wpa}.

We consider proton--proton collisions with $\sqrt{s}=7, 8, 13$ and $14\,\TeV$.
As far as EW couplings are concerned, we use the so-called $G_\mu$ scheme,
where the input parameters are $G_F$, $m_W$, $m_Z$.
More precisely and consistent with the \OpenLoops{} implementation, we use the complex \w{} 
and \z{} boson masses
to define the EW mixing angle as 
$\cos\theta_W^2=(m_W^2-i\Gamma_W\,m_W)/(m_Z^2-i\Gamma_Z\,m_Z)$,
and set $G_F = 1.16639\times 10^{-5}$~GeV$^{-2}$, $m_W=80.385$ GeV, $\Gamma_W=2.0854$ GeV,
$m_Z = 91.1876$~GeV, $\Gamma_Z=2.4952$ GeV. With these inputs, the relevant 
leading-order branching fractions are \mbox{$BR(W^\pm\to\nu\ell^\pm)=0.108984$} 
and \mbox{$BR(Z\to\ell^+\ell^-)=0.0336313$}.
We set the CKM matrix to unity\footnote{The numerical effect of the CKM matrix up to NLO 
is to reduce the cross section by less than $1\%$.}.
We employ the NNPDF3.0~\cite{Ball:2014uwa} sets of parton distributions with 
$\as(m_Z)=0.118$, and the $\as$ running is evaluated at each corresponding order
(i.e., $(n+1)$-loop $\as$ running at N$^n$LO, with $n=0,1,2$).
We consider $N_f=5$ massless quark flavours. The central
renormalization ($\mu_R$) and factorization ($\mu_F$) scales are set to
$\mu_R=\mu_F=\mu_0\equiv\frac{1}{2}(m_Z+m_W)=85.7863\,\GeV$. 
Scale uncertainties are computed by the customary $7$-point variation, i.e., we vary 
independently $0.5\mu_0\le \mu_R,\mu_F\le 2 \mu_0$ with the constraint
$0.5\le \mu_R/\mu_F\le 2$. If not stated otherwise, all cross sections presented 
in the following are summed over the electrical charges of the final-state \w{} bosons, 
i.e.\ they refer to $\sigma(pp\to W^+Z)+\sigma(pp\to W^-Z)$.

\renewcommand{\baselinestretch}{1.5}
\input{tables/XS_ratios_onshell.tex}
\renewcommand{\baselinestretch}{1.0}

\begin{figure}[t]
\centering
\includegraphics[width=0.7\textwidth]{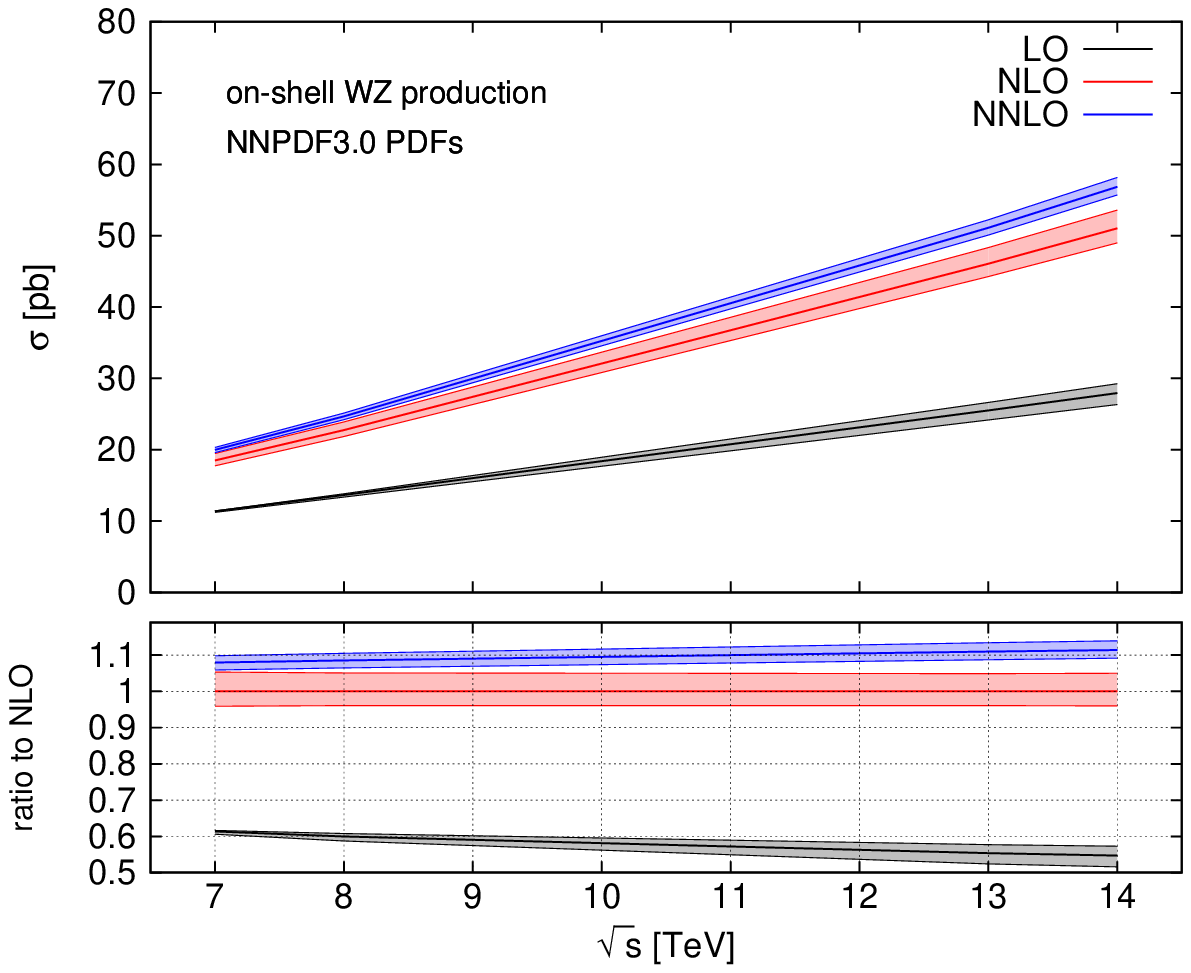}
\caption{On-shell \wz{} cross section as a function of the \cme{}
at \lo{}, \nlo{} and \nnlo{}. In the lower panel the curves of the main frame are
normalized to the central \nlo{} prediction. The bands correspond to scale variations 
as described in the text.}
\label{fig:onshell}
\end{figure}

We start the presentation of our results by considering the on-shell \wz{} 
cross sections\footnote{In this on-shell computation, the gauge-boson widths are set 
to zero, and a real EW mixing angle is used correspondingly.}.
In \refta{tab:onshell} we report the LO, NLO and NNLO cross sections and 
scale uncertainties in the $\sqrt{s}$ range from 7 to 14\,TeV.
The same results are shown in \reffi{fig:onshell}.
The main interest of these on-shell results is that they can be unambiguously defined 
without choosing a specific mass window for the \z{} boson decay products.
Consistent with \citere{Campbell:2011bn}, the \nlo{} corrections
are quite large and increase the \lo{} result
by $63\%$ to $83\%$ for \cmes{} between $7$ and $14$\,TeV. 
The \nnlo{} corrections further increase the \nlo{} results, 
and the effect ranges from $8\%$ to $11\%$.
We note that in contrast to \zz{} and \ww{} production, these are purely 
genuine \nnlo{} corrections to the $q\bar{q}$ channel:
As the Born-level final state is electrically charged, the production process does not 
receive contributions from a loop-induced gluon-fusion channel.
Due to the absence of such a loop-induced gluon--gluon channel, which usually features 
large corrections, \nlo{} scale variations might be expected to
give a reliable estimate of possible effects
at the \nnlo{} and beyond. However, this turns out to be not the case in general: 
In particular at large collider energies, the \nnlo{} corrections are roughly twice 
as large as the uncertainties estimated by scale variations at \nlo{}.
We note that the scale uncertainties drop from about $\pm 5\%$ at \nlo{} to 
about $\pm 2\%$ at \nnlo{}.

Similarly to what happens in the case of $W\gamma$ production~\cite{Grazzini:2015nwa},
the rather large impact of radiative corrections is due to the existence of a 
\textit{radiation zero} in the Born scattering amplitudes. More precisely,
the partonic $W\gamma$ tree amplitude exhibits an \textit{exact} radiation zero 
at $\theta^*=1/3$, where $\theta^*$ is the scattering angle 
in the centre-of-mass frame~\cite{Mikaelian:1979nr}.
Analogously, the partonic on-shell Born \wz{} amplitude exhibits an 
\textit{approximate} radiation zero~\cite{Baur:1994ia}\footnote{The approximate nature 
of the radiation zero for \wz{} production is due to the fact that it appears
only in the dominant helicity amplitudes for this process~\cite{Baur:1994ia}.
On the contrary, the $W\gamma$ process has an exact zero in all the helicity amplitudes, 
due to the presence of the massless photon.}.
The radiation zero is broken by real corrections starting from the \nlo{}, but 
suppresses the LO cross section, thus leading to an increased impact of higher-order 
corrections.

For completeness, in \refta{tab:onshellcharges} we provide cross sections and 
relative corrections for the two contributing processes $pp\to W^+ Z$ and $pp\to W^-Z$. 
The \wpz{} rate, being driven by $u{\bar d}$ scattering, is larger than the \wmz{} rate, 
which is driven by $d{\bar u}$ scattering. 
The difference decreases as $\sqrt{s}$ increases.
As expected, radiative corrections to the two processes are very similar. 
They turn out to be slightly larger for \wmz{} than for \wpz{}, leading to a reduction 
of the $\sigma^{\wpz}/\sigma^{\wmz}$ ratio at higher perturbative orders. 
This difference in the ratios, however, is due to differences in the PDFs used at 
each order, and it decreases with increasing collider energies, 
never exceeding $1\%$ at the \nnlo{}.

\renewcommand{\baselinestretch}{1.5}
\input{tables/XS_ratios_onshell_charges.tex}
\renewcommand{\baselinestretch}{1.0}

From now on, all our results contain the full off-shell effects.
The ATLAS and CMS collaborations report inclusive \wz{} cross sections obtained by 
considering a mass window on the reconstructed \z{} boson.
The mass window slightly  differs between ATLAS and CMS: 
While ATLAS uses $66\,\GeV<m(\z{})<116\,\GeV$ for their measurements at $7$ and $8$\,TeV 
(a measurement at $13\,\TeV$ has not been published so far), CMS applies 
a cut of $71\,\GeV<m(\z{})<111\,\GeV$ for their measurements at $7$ and $8$\,TeV and
$60\,\GeV<m(\z{})<120\,\GeV$ for their measurement at $13\,\TeV$. 
We find that the numerical differences between the cross sections computed 
in the different mass windows are at the 1\% level,
and thus significantly smaller than the current experimental uncertainties.
When considering the relative effects of radiative corrections, the impact of the 
different mass windows is completely negligible. 
Nevertheless, we will consistently apply the respective mass windows when comparing 
to data in the following.

We first present results for the ATLAS definition of the \wz{} cross sections, 
reported in \refta{tableatlas}, where we compare with the $7$ and $8$\,TeV ATLAS 
measurements of \citere{Aad:2012twa} and \citere{Aad:2016ett}, respectively. 
Comparing these cross sections in absolute terms to the on-shell case, 
we find a reduction by roughly 3\% due to the applied mass-window cut and 
genuine off-shell effects; however, as anticipated, the relative impact of 
radiative corrections remains widely unchanged, 
again ranging between $63\%$ and $83\%$ at NLO and between $8\%$ and $11\%$ at NNLO 
for the collider energies under consideration. 
Also the scale uncertainty bands stay almost identical when including 
off-shell effects and applying the ATLAS mass cut.

Comparing with the experimentally measured cross sections 
from \citeres{Aad:2012twa,Aad:2016ett}, we find that the inclusion of \nnlo{} corrections 
clearly improves the agreement between data and theory, in particular at $8\,\TeV$, 
where the measurement is most precise.
While the central \nlo{} prediction is roughly $2\sigma$ away from the measured 
cross section at $8\,\TeV$, the \nnlo{} prediction is right on top of the data 
with fully overlapping uncertainty bands.

\renewcommand{\baselinestretch}{1.5}
\input{tables/XS_ATLAS.tex}
\renewcommand{\baselinestretch}{1.0}
\renewcommand{\baselinestretch}{1.5}
\input{tables/XS_CMS.tex}
\renewcommand{\baselinestretch}{1.0}

Next, we provide theory predictions for the \wz{} cross sections as defined by CMS 
in \refta{tablecms}, where we also quote the results of the CMS measurements 
performed at $7$ and $8\,\TeV$ (reported in \citere{CMS:2013qea}), 
and at $13\,\TeV$ (reported in \citere{CMS:2016kma}).
As already anticipated, the precise definition of the \z{}-mass window has only a 
very mild impact on the cross section. 
In particular, both the relative size of higher-order corrections and the bands obtained 
by scale variation are almost identical to the ones obtained with the ATLAS definition. 
Comparing with the measured cross sections, we again find excellent agreement 
between our \nnlo{} predictions and the cross sections reported by CMS for 
$\sqrt{s}=7$ and $8\,\TeV$, where the inclusion of \nnlo{} corrections again 
clearly improves the agreement, in particular at $8\,\TeV$. 
The measurement at $13\,\TeV$ undershoots the \nnlo{} prediction, being consistent 
with it only at the level of about $2\sigma$. 
However, at this early stage of LHC Run 2, the measurement still comes with quite 
large experimental uncertainties of both systematical and statistical nature, and 
the measured cross section might still be subject to a significant shift 
with respect to its central value, once statistics increases.

\begin{figure}[tpb]
\centering
\includegraphics[width=0.8\textwidth]{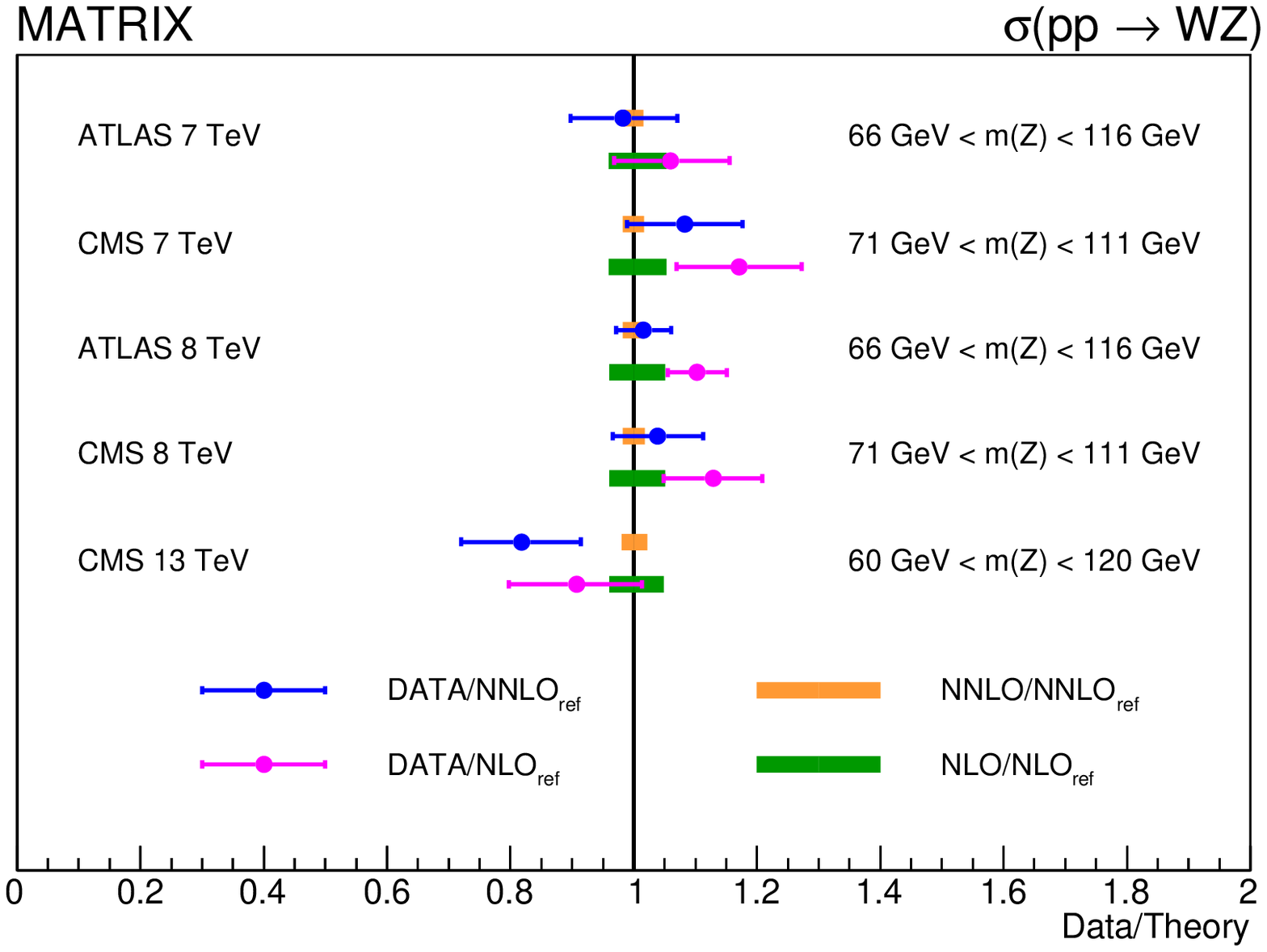}
\caption{Summary plot for comparison of NLO and NNLO predictions with the available 
LHC measurements of the total \wz{} cross section. Theory uncertainties are obtained 
through scale variations as described in the text.}
\label{fig:summary}
\end{figure}

\reffi{fig:summary} shows a summary plot where we compare NLO and NNLO predictions 
to all available LHC measurements of the total \wz{} cross section.

We have presented the first exact NNLO QCD computation for the production 
of \wz{} pairs at the LHC.
We have considered both the case in which the vector bosons are on shell,
and the case in which off-shell effects are accounted for.
The \nnlo{} corrections are sizeable and range from about $8\%$ at $\sqrt{s}=7$\,TeV to 
about $11\%$ at $\sqrt{s}=14$\,TeV with respect to the \nlo{} prediction, 
significantly exceeding what might be expected from \nlo{} scale variations. 
The remaining scale uncertainties at \nnlo{} are at the level of about $2\%$.
We have stressed that the large size of QCD radiative corrections is due to 
an approximate radiation zero which is present in the on-shell amplitude at LO. 
Nonetheless, since all partonic channels are included at NNLO, 
we expect that scale variations should provide the correct order of magnitude 
of the uncertainty from yet uncalculated higher-order QCD corrections beyond NNLO.

When off-shell effects for the $W$ and $Z$ bosons are accounted for,
our results can be compared to the inclusive cross sections
presented by ATLAS and CMS, provided the same range in the virtuality
of the $Z$ boson is imposed.
We find that the inclusion of \nnlo{} corrections is mandatory
in order to obtain agreement (within $1\sigma$) with the inclusive cross sections 
measured by ATLAS and CMS in Run 1 of the LHC. 
The computed corrections will be even more important at $13\,\TeV$, once enough 
statistics is accumulated. 
Since our computation already involves the full helicity amplitudes and all 
off-shell effects, it can be extended to include realistic selection cuts 
on the final-state leptons and to provide predictions
in the fiducial volume in which the measurements are carried out.
It will also be possible to provide precise background predictions in new-physics searches 
based on the trilepton + missing energy signature.

Our calculation was performed with the numerical program \Matrix{}, which is able to 
carry out fully-exclusive \nnlo{} computations for a wide class of processes 
at hadron colliders. 
We are planning further applications of our framework to other important LHC processes.

\noindent {\bf Acknowledgements.}
This research was supported in part by the Swiss National Science Foundation (SNF) under 
contracts CRSII2-141847, 200021-156585, by 
the Research Executive Agency (REA) of the European Union under the Grant Agreement 
number PITN--GA--2012--316704 ({\it HiggsTools}), and by the National Science 
Foundation under Grant No. NSF PHY11-25915.

\end{document}

%% file: tables/XS_ratios_onshell.tex
\begin{table}[t]
\begin{center}
\begin{tabular}{|c| c| c| c| c| c|}
\hline
$\sqrt{s}$ & $\sigma_{\textrm{LO}}$ [pb] & $\sigma_{\textrm{NLO}}$ [pb] & $\sigma_{\textrm{NNLO}}$ [pb] & $\sigma_{\textrm{NLO}}/\sigma_{\textrm{LO}}$ & $\sigma_{\textrm{NNLO}}/\sigma_{\textrm{NLO}}$\\ [0.5ex]
\hline
\Tstrut
$7$ 
& $11.354(1)\,^{+0.5\%}_{-1.2\%}$ 
& $18.500(1)\,^{+5.3\%}_{-4.1\%}$ 
& $19.973(13)\,^{+1.7\%}_{-1.9\%}$ 
& $+62.9\%$ 
& $+\phantom{0}8.0\%$ 
\Bstrut\\
\Tstrut
$8$ 
& $13.654(1)\,^{+1.3\%}_{-2.1\%}$ 
& $22.750(2)\,^{+5.1\%}_{-3.9\%}$ 
& $24.690(16)\,^{+1.8\%}_{-1.9\%}$ 
& $+66.6\%$
& $+\phantom{0}8.5\%$
\Bstrut\\
\Tstrut
$13$ 
& $25.517(2)\,^{+4.3\%}_{-5.3\%}$ 
& $46.068(3)\,^{+4.9\%}_{-3.9\%}$ 
& $51.11(3)\phantom{00}\,^{+2.2\%}_{-2.0\%}$ 
& $+80.5\%$
& $+10.9\%$
\Bstrut\\
\Tstrut
$14$ 
& $27.933(2)\,^{+4.7\%}_{-5.7\%}$ 
& $51.038(3)\,^{+5.0\%}_{-4.0\%}$ 
& $56.85(3)\phantom{00}\,^{+2.3\%}_{-2.0\%}$ 
& $+82.7\%$
& $+11.4\%$
\Bstrut\\
\hline
\end{tabular}
\end{center}
\renewcommand{\baselinestretch}{1.0}
\caption{\label{tab:onshell} Total on-shell \wz{} cross sections at \lo{}, \nlo{} and \nnlo{} for relevant  collider energies; the last two columns contain the relative corrections at NLO and NNLO, respectively.}
\end{table}

%% file: tables/XS_ratios_onshell_charges.tex
\begin{table}[t]
\begin{center}
\begin{tabular}{|c| c| c| c| c| c| c|}
\hline
$\sqrt{s}$ 
& process
& $\sigma_{\textrm{LO}}$ [pb] 
& $\sigma_{\textrm{NLO}}$ [pb] 
& $\sigma_{\textrm{NNLO}}$ [pb]
& $\sigma_{\textrm{NLO}}/\sigma_{\textrm{LO}}$
& $\sigma_{\textrm{NNLO}}/\sigma_{\textrm{NLO}}$
\\ [0.5ex]
\hline
\Tstrut
$7$ 
& \wpz
& $\phantom{0}7.343(1)\,^{+0.4\%}_{-1.1\%}$ 
& $11.867(1)\,^{+5.3\%}_{-4.1\%}$ 
& $12.790(11)\,^{+1.8\%}_{-1.9\%}$ 
& $+61.6\%$
& $+\phantom{0}7.8\%$
\Bstrut\\
\Tstrut

& \wmz
& $\phantom{0}4.011(1)\,^{+0.7\%}_{-1.4\%}$ 
& $\phantom{0}6.633(1)\,^{+5.4\%}_{-4.1\%}$ 
& $\phantom{0}7.183(6)\phantom{0}\,^{+1.7\%}_{-1.9\%}$ 
& $+65.4\%$
& $+\phantom{0}8.3\%$
\Bstrut\\
\hline
\Tstrut
$8$ 
& \wpz
& $\phantom{0}8.741(1)\,^{+1.2\%}_{-2.0\%}$ 
& $14.443(1)\,^{+5.0\%}_{-3.9\%}$ 
& $15.650(14)\,^{+1.9\%}_{-1.9\%}$ 
& $+65.2\%$
& $+\phantom{0}8.4\%$
\Bstrut\\
\Tstrut

& \wmz
& $\phantom{0}4.913(1)\,^{+1.5\%}_{-2.3\%}$ 
& $\phantom{0}8.307(1)\,^{+5.1\%}_{-3.9\%}$ 
& $\phantom{0}9.040(8)\phantom{0}\,^{+1.8\%}_{-1.8\%}$ 
& $+69.1\%$
& $+\phantom{0}8.8\%$
\Bstrut\\
\hline
\Tstrut
$13$ 
& \wpz
& $15.787(2)\,^{+4.1\%}_{-5.1\%}$ 
& $28.251(3)\,^{+4.9\%}_{-3.9\%}$ 
& $31.33(3)\phantom{00}\,^{+2.3\%}_{-2.0\%}$ 
& $+79.0\%$
& $+10.9\%$
\Bstrut\\
\Tstrut

& \wmz
& $\phantom{0}9.730(1)\,^{+4.5\%}_{-5.5\%}$ 
& $17.817(2)\,^{+4.9\%}_{-4.0\%}$ 
& $19.78(2)\phantom{00}\,^{+2.2\%}_{-2.0\%}$ 
& $+83.1\%$
& $+11.0\%$
\Bstrut\\
\hline
\Tstrut
$14$ 
& \wpz
& $17.199(2)\,^{+4.6\%}_{-5.6\%}$ 
& $31.147(3)\,^{+4.9\%}_{-4.0\%}$ 
& $34.68(3)\phantom{00}\,^{+2.4\%}_{-2.1\%}$ 
& $+81.1\%$
& $+11.3\%$
\Bstrut\\
\Tstrut

& \wmz
& $10.733(1)\,^{+4.9\%}_{-6.0\%}$ 
& $19.891(2)\,^{+5.0\%}_{-4.0\%}$ 
& $22.17(2)\phantom{00}\,^{+2.2\%}_{-2.0\%}$ 
& $+85.3\%$
& $+11.4\%$
\Bstrut\\
\hline
\end{tabular}
\end{center}
\renewcommand{\baselinestretch}{1.0}
\caption{\label{tab:onshellcharges} Total on-shell \wz{} cross sections at \lo{}, \nlo{} and \nnlo{}, together with the relative corrections to the respective lower order, for relevant collider energies, separated according to the charge of the final states.}
\end{table}

%% file: tables/XS_ATLAS.tex
\begin{table}[t]
\begin{center}
\begin{tabular}{|c| c| c| c| l|}
\hline
$\sqrt{s}$ & $\sigma_{\textrm{LO}}$ [pb] & $\sigma_{\textrm{NLO}}$ [pb] & $\sigma_{\textrm{NNLO}}$ [pb] & \multicolumn{1}{c|}{$\sigma_{\textrm{ATLAS}}$ [pb]} \\ [0.5ex]
\hline
\Tstrut
$7$ & $11.028(8)^{+0.5\%}_{-1.2\%}$ & $17.93(1)^{+5.3\%}_{-4.1\%}$ & $19.34(3)^{+1.6\%}_{-1.8\%}$ & $19.0\,^{+1.4}_{-1.3}{\rm (stat)}^{+0.9}_{-0.9}{\rm (syst)}^{+0.4}_{-0.4}{\rm (lumi)}\phantom{^{+0.0}_{-0.0}{\rm (th)}}$\Bstrut\\
\Tstrut
$8$ & $13.261(9)^{+1.3\%}_{-2.1\%}$ & $22.03(2)^{+5.1\%}_{-3.9\%}$ & $23.92(3)^{+1.7\%}_{-1.8\%}$ & $24.3\,^{+0.6}_{-0.6}{\rm (stat)}^{+0.6}_{-0.6}{\rm (syst)}^{+0.5}_{-0.5}{\rm (lumi)}^{+0.4}_{-0.4}{\rm (th)}$\Bstrut\\
\Tstrut
$13$ & $24.79(2)\phantom{0}^{+4.2\%}_{-5.2\%}$ & $44.67(3)^{+4.9\%}_{-3.9\%}$ & $49.62(6)^{+2.2\%}_{-2.0\%}$ & \Bstrut\\
\Tstrut
$14$ & $27.14(2)\phantom{0}^{+4.7\%}_{-5.7\%}$ & $49.50(3)^{+4.9\%}_{-4.0\%}$ & $55.10(7)^{+2.3\%}_{-2.0\%}$ & \Bstrut\\
\hline
\end{tabular}
\end{center}
\renewcommand{\baselinestretch}{1.0}
\caption{\label{tableatlas} Total cross sections with ATLAS mass window $66\,\GeV<m(\z{})<116\,\GeV$ 
at \lo{}, \nlo{} and \nnlo{}. The available ATLAS data from \citeres{Aad:2012twa,Aad:2016ett} are also shown.}

%% file: tables/XS_CMS.tex
\begin{center}
\begin{tabular}{|c| c| c| c| l|}
\hline
$\sqrt{s}$ & $\sigma_{\textrm{LO}}$ [pb] & $\sigma_{\textrm{NLO}}$ [pb] & $\sigma_{\textrm{NNLO}}$ [pb] & \multicolumn{1}{c|}{$\sigma_{\textrm{CMS}}$ [pb]} \\ [0.5ex]
\hline
\Tstrut
$7$ & $10.902(7)^{+0.5\%}_{-1.2\%}$ & $17.72(1)^{+5.3\%}_{-4.1\%}$ & $19.18(3)^{+1.7\%}_{-1.8\%}$ & $20.76\,^{+1.32}_{-1.32}{\rm (stat)}^{+1.13}_{-1.13}{\rm (syst)}^{+0.46}_{-0.46}{\rm (lumi)}$\Bstrut\\
\Tstrut
$8$ & $13.115(9)^{+1.3\%}_{-2.1\%}$ & $21.80(2)^{+5.1\%}_{-3.9\%}$ & $23.68(3)^{+1.8\%}_{-1.8\%}$ & $24.61\,^{+0.76}_{-0.76}{\rm (stat)}^{+1.13}_{-1.13}{\rm (syst)}^{+1.08}_{-1.08}{\rm (lumi)}$\Bstrut\\
\Tstrut
$13$ & $25.04(2)\phantom{0}^{+4.3\%}_{-5.3\%}$ & $45.09(3)^{+4.9\%}_{-3.9\%}$ & $49.98(6)^{+2.2\%}_{-2.0\%}$ & $40.9\phantom{0}\,^{+3.4}_{-3.4}{\rm (stat)}^{+3.1}_{-3.3}{\rm (syst)}^{+1.3}_{-1.3}{\rm (lumi)}^{+0.4}_{-0.4}{\rm (th)}$\Bstrut\\
\Tstrut
$14$ & $27.39(2)\phantom{0}^{+4.7\%}_{-5.7\%}$ & $49.91(4)^{+4.9\%}_{-4.0\%}$ & $55.60(7)^{+2.3\%}_{-2.0\%}$ & \Bstrut\\
\hline
\end{tabular}
\end{center}
\renewcommand{\baselinestretch}{1.0}
\caption{\label{tablecms} Total cross sections with CMS mass windows of $71\,\GeV<m(\z{})<111\,\GeV$ for $7$ and $8\,\TeV$,
and $60\,\GeV<m(\z{})<120\,\GeV$ for $13$ and $14\,\TeV$,
at \lo{}, \nlo{} and \nnlo{}. The available CMS data from \citeres{CMS:2013qea,CMS:2016kma} are also shown.}
\end{table}

%% file: wznnlo.bbl
\begin{thebibliography}{99}
\bibitem{Kallweit:Munich}
S.~Kallweit,
\Munich{} is the abbreviation of ``MUlti-chaNnel Integrator at Swiss~(CH) precision''---an automated parton level NLO 
generator. 
In preparation.
\end{thebibliography}
